\documentclass[]{aa}
\usepackage{psfig}
\begin{document}

\title{Nova Cygni 2001/2 = V2275~Cyg\thanks{Based on the data
obtained at the David Dunlap Observatory, University of Toronto}}

\author{L.L. Kiss\inst{1} \and N. G\H{o}gh\inst{1} \and
J. Vink\'o\inst{2} \and G. F\H{u}r\'esz\inst{1} \and B. Cs\'ak\inst{1}
\and H. DeBond\inst{3} \and
J.R. Thomson\inst{3} \and A. Derekas\inst{1}}

\institute{Department of Experimental Physics and Astronomical Observatory,
University of Szeged,
Szeged, D\'om t\'er 9., H-6720 Hungary \and
Department of Optics and Quantum Electronics, University of Szeged,
POB 406, Szeged, H-6701 Hungary \and
David Dunlap Observatory, University of Toronto, Richmond Hill, Canada}

\titlerunning{Optical spectroscopy of V2275~Cyg}
\authorrunning{Kiss et al.}
\offprints{l.kiss@physx.u-szeged.hu}
\date{}

\abstract{
We present an analysis of low- and medium resolution spectra of the
very fast nova, Nova Cygni 2001/2 (V2275~Cyg) obtained at nine epochs in 
August, September and October, 2001. The expansion velocity from
hydrogen Balmer lines is found to be 2100 km~s$^{-1}$, although 
early H$\alpha$ profile showed a weak feature at $-$3500 km~s$^{-1}$, too.
The overall appearance of the optical spectrum is dominated by 
broad lines of H, He and N, therefore, the 
star belongs to the ``He/N'' subclass of novae defined
by Williams (1992). Interstellar lines and bands, as well as $BV$ photometry
taken from the literature yielded to a fairly high reddening 
of $E(B-V)=1\fm0\pm0\fm1$. The visual light curve was used to deduce 
M$_{\rm V}$ by the maximum magnitude versus rate of decline 
relationship. The resulting parameters are: $t_0=2452141.4^{+0.1}_{-0.5}$,
$t_2=2.9\pm0.5$ days, $t_3=7\pm1$ days, M$_{\rm V}=-9\fm7\pm0\fm7$.
Adopting these parameters, the star lies between 3 kpc and 8 kpc from
the Sun.  
}
 
\maketitle

\section{Introduction}

Nova Cygni 2001/2 was discovered by A. Tago on two T-Max 400 films
taken on Aug. 18 at 8\fm8 apparent brightness. One day before 
discovery, nothing was visible at the nova location down to 
12 mag (Nakamura et al. 2001).
The spectroscopic confirmation was given by subsequent optical 
spectroscopy revealing hydrogen Balmer emission lines with 
deep P Cygni profiles. The H$\alpha$ line profile suggested 
an early expansion velocity of 1700 km~s$^{-1}$ (Ayani 2001).
The nova was also discovered independently by K. Hatayama (Nakano et al.
2001). A possible progenitor with USNO red magnitude 18\fm8 and blue 
mag magnitude 19\fm6 was identified by P. Schmeer (Schmeer et al. 2001).  

Early photometric data consist of photographic and CCD photometric data 
published in the IAU Circulars. The full light curve can be reconstructed
using visual data collected by the VSNET group\footnote{\tt
http://www.kusastro.kyoto-u.ac.jp/vsnet}. The visual maximum occured at
6\fm8 on 2001 Aug. 19.9 UT. Simultaneous color measurements
were published by Sostero \& Leopardo (in Nakano et al. 2001), who gave 
$B-V$=1\fm1 in the maximum, suggesting substantial reddening. 
The color remained around $B-V\sim 1\fm0$  during the first week after
the maximum, when the apparent brightness decreased down to 9\fm2
(Samus et al. 2001).

The main aim of this paper is to present optical and far red spectra taken
after the maximum, between $\Delta$t = +2.3 d to
$\Delta$t = +59.1 d. The low- and medium resolution spectra
were used to determine the main outburst properties, the
expansion velocity and the interstellar
reddening. We also make use of all publicly available visual
photometric data collected by
the VSNET group to estime the rates of decline and to check the photometric
phases of the obtained spectra.

\section{Observations}

The spectroscopic observations were carried out with the
Cassegrain-spectrograph attached to the 1.88-m telescope of the David Dunlap
Observatory (Richmond Hill, Canada). The spectra were obtained on nine
nights between August and October, 2001. The detector was a Thomson
1024$\times$1024 CCD chip (with a 6 e$^-$ readout noise). The slit
width was 303$\mu$ corresponding to 1\farcs8 on the sky. As the
typical observing circumstances at DDO are non-photometric, we did
not attempt to flux calibrate the data. All spectra presented 
throughout the paper were continuum normalized, with few exceptions
where the normalization was impossible due to the short wavelength span.
Further observational details (the gratings, wavelength range, 
photometric phase) are given in Table\ 1. The same instrument and 
setup was used earlier for spectroscopic investigation of two
recent novae (V1494~Aql - Kiss \& Thomson 2000; CI~Aql - Kiss et al. 2001).  

\begin{table}
\caption{Observing log (MJD=HJD$-$2452000). $\Delta t$ is the time
after visual maximum, estimated from the visual light curve (see Sect.\ 5).}
\begin{center}
\begin{tabular}{llllrr}
\hline
Date & MJD & grating & range & ${\lambda \over \Delta \lambda}$ & $\Delta t$\\
2001 &     & ln/mm   & (\AA) &                            & (d) \\
\hline
Aug. 22 & 143.70 & 831 & \bf 6300--6800 & 6000 & +2.3\\
        & 143.74 & 600 & 3900--4500 & 7000 & +2.3\\ 
Sep. 5  & 157.85 & 100 & 4200--7900 & 1400 & +16.5\\
        & 157.87 & 1800 & 5800--6000 & 9800 & +16.5\\
Sep. 7  & 159.57 & 1800 & 6500--6700 & 11000 & +18.2\\
Sep. 9  & 161.64 & 1800 & 6500--6700 & 11000 & +20.2\\
Sep. 10 & 162.53 & 600 & 8350--8950 &  11000 & +21.1\\
Sep. 12 & 164.57 & 831 & 6400--6900 &  6000 & +23.2\\
Oct. 8  & 191.56 & 100 & 4300--8000 & 1400 & +50.2\\
        & 191.66 & 831 & 6300--6800 & 6000 & +50.3\\
        & 191.72 & 1800 & 6450--6650 & 11000 & +50.3\\
Oct. 15 & 198.53 & 1800 & 6450--6650 & 11000 & +57.1\\
Oct. 17 & 200.53 & 1800 & 6450--6650 & 11000 & +59.1\\
\hline
\end{tabular}
\end{center}
\end{table}

The spectra were reduced with standard IRAF tasks, including bias
removal, flat-fielding, aperture extraction (with the task {\it doslit})
and wavelength calibration. For the latter, FeAr, ThAr and FeNe spectral
lamp exposures were used (depending on the actual wavelength range), which
were taken immediately before and after every stellar exposures. The
integration times varied between 5 and 30 minutes, according to the 
apparent brightness, wavelength range and resolving power.

\section{Description of the spectra}

The first spectroscopic measurements were carried out on Aug. 22,
approximately 2 days after the visual maximum. We recorded a 600 \AA\ long
spectrum in the blue region and an H$\alpha$ profile. 
The blue spectrum showed  the prominent hydrogen Balmer series
lines (H$\gamma$, H$\delta$, H$\epsilon$) and two broad iron lines. Because
of the presence of overlapping broad emission lines, the continuum level is
uncertain. Therefore, this spectrum could not be normalized. It is plotted in
Fig.\ 1, where the individual hydrogen line profiles (both H$\alpha$ and
those from the blue spectrum) are shown in the bottom panel. 
The P Cygni absorption is very similar in all four lines suggesting expansion
velocity about 2100 km~s$^{-1}$, though a weak feature in the H$\alpha$ is
also visible at $\sim-$3500 km~s$^{-1}$ (the velocity resolution is 
about 50 km~s$^{-1}$).

\begin{figure}
\begin{center}
\leavevmode
\psfig{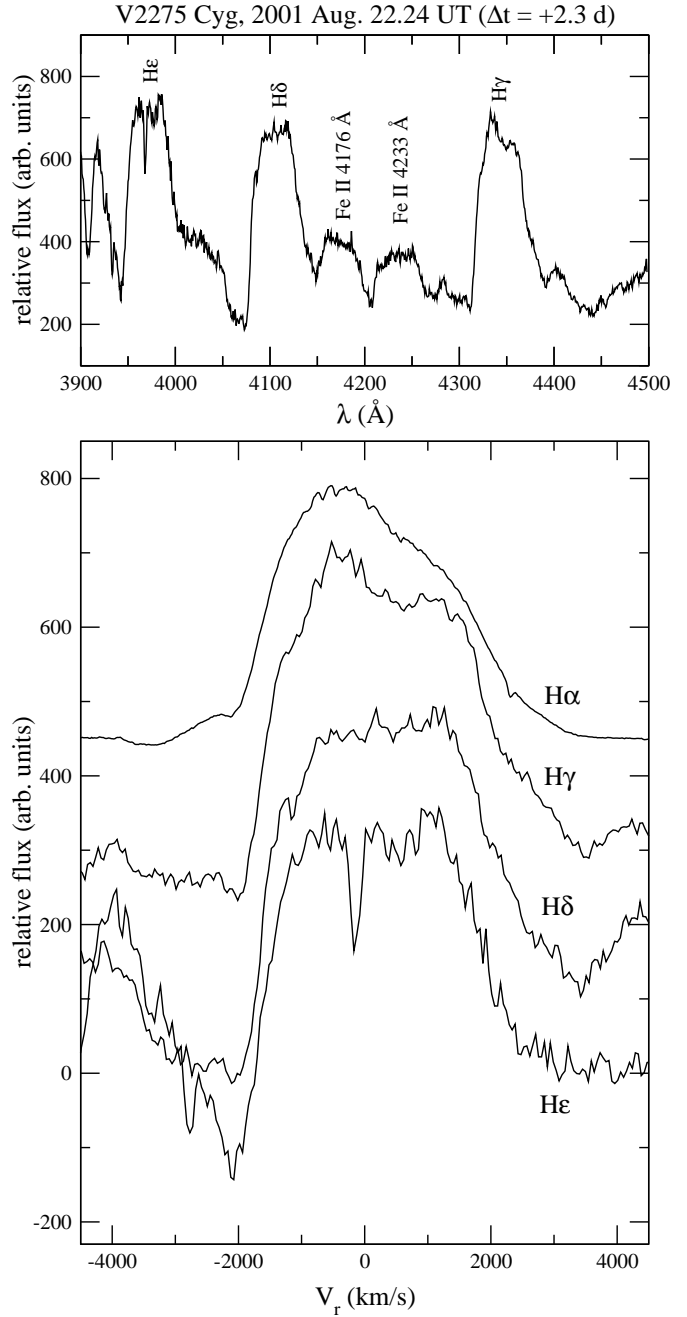}
\caption{{\it Top}: The intermediate resolution spectrum of V2275~Cyg
in the blue region obtained shortly after the visual maximum. {\it Bottom}:
The velocity structure of hydrogen line profiles. The well-defined
P Cygni profiles suggest an expansion velocity of 2100 km~s$^{-1}$.}
\end{center}
\label{f1}
\end{figure}

\begin{figure}
\begin{center}
\leavevmode
\psfig{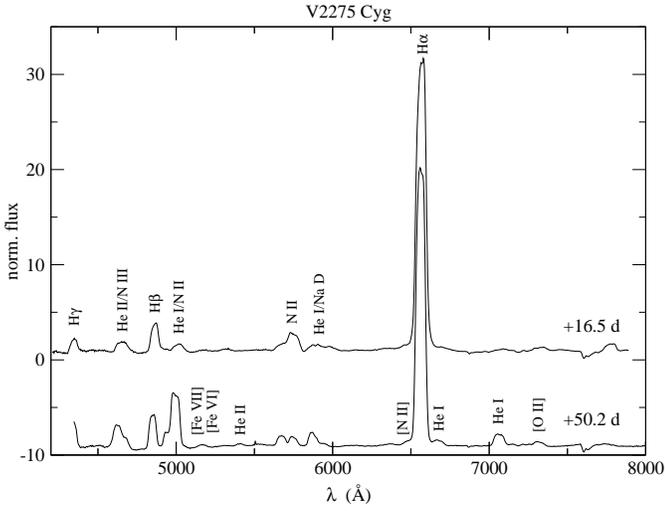}
\caption{Low-resolution spectra of V2275~Cyg}
\end{center}
\label{f2}
\end{figure}

\begin{figure}
\begin{center}
\leavevmode
\psfig{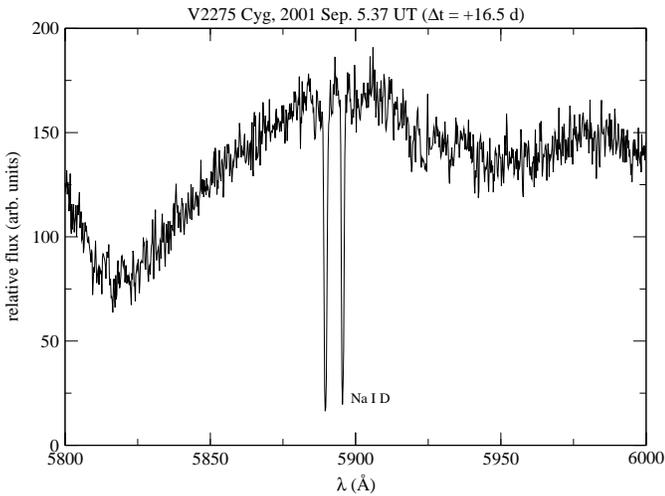}
\caption{The He I 5876 and Na I D blend. Note the strong interstellar 
component of the sodium doublet implying fairly high reddening}
\end{center}
\label{f3}
\end{figure}

The next run was on Sep. 5, when the whole optical region between 4200 \AA\
and 7900 \AA\ was recorded together with a medium-resolution spectrum of the
He I 5876/Na I D blend. Further low-resolution spectrum was obtained on Oct.
8 providing a rough picture of the changing spectral appearence after 
maximum (see Fig.\ 2). Using the extensive observational material published
by Williams and his co-workers (Williams et al. 1991, Williams 1992,
Williams et al. 1994) we could identify the following lines/blends: hydrogen
Balmer series from H$\alpha$ to H$\gamma$, N III 4640/He II 4686, N II
5001/He I 5016, N II 5679, He I 5876/Na I D. The later spectrum showed
significant differences in strength of lines and some additional emission
features appeared: [Fe VII] 5159 + [Fe VI 5176], He II 5412, [N II] blended
with H$\alpha$, He I 6678, He I 7065, [O II] 7325. The single Na I D
observation revealed very strong interstellar component (Fig.\ 3), which
can be used as a reddening indicator (Sect.\ 4). Here the
unknown  continuum level again disables the continuum normalization 

The largest number of spectra addressed the evolution of the H$\alpha$ line.
Being the strongest emission line, it could be well observed even 
at later phases in October, 2001, when the apparent magnitude decreased below
12\fm0.  Fig.\ 4 summarizes the observed line profile variations. As in the
case of CI~Aql (Kiss et al. 2001), we detected the strong diffuse
interstellar band (DIB) at 6613 \AA. The general appearence of the line
profile remained essentially the same, only the maximum of 
the continuum normalized flux
changed between 15 and 35. We note the remarkably stable system of
narrow absorption and emission features which can be identified in every
H$\alpha$ profiles. The existence of this system is supported by the single
far red spectrum (8350 \AA-8950 \AA) obtained on Sep. 10. It covers the
prominent O I 8446 emission line showing exactly the same system of narrow
components (see Fig.\ 4). We have determined the component radial
velocities, though it is difficult to judge which features are in absorption
and which are in emission. The radial velocities 
of the marked features are presented in Table\ 2 (the velocity resolution
of spectra shown in Fig.\ 4 is between 20 and 60 km~s$^{-1}$, depending
on the actual spectral resolution -- see Table 1).

\begin{figure}
\begin{center}
\leavevmode
\psfig{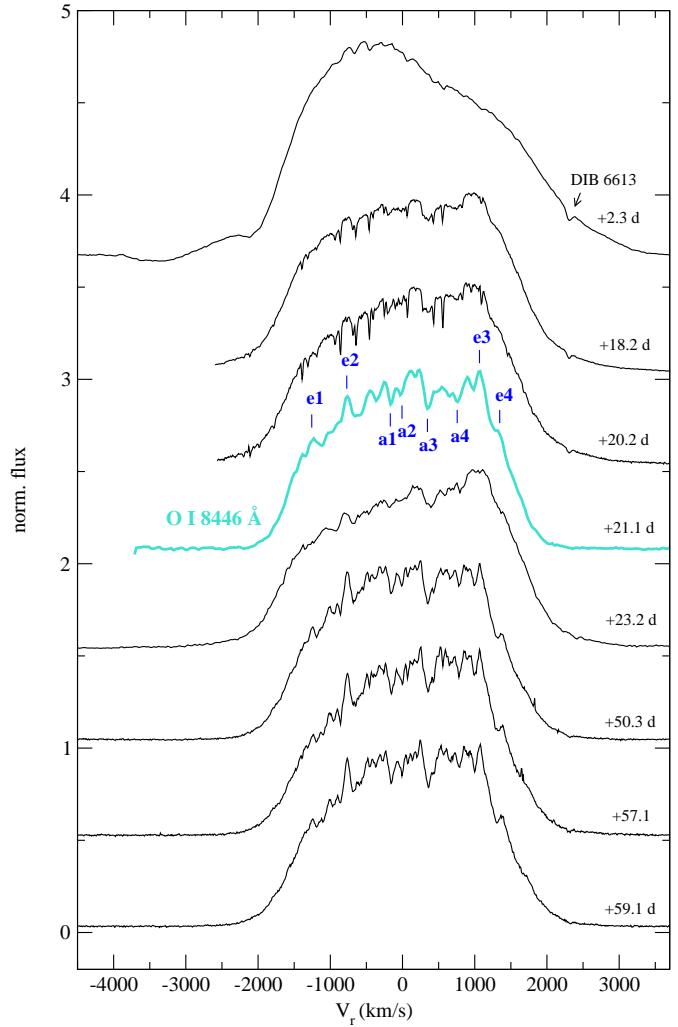}
\caption{The evolution of the H$\alpha$ profile. The fluxes at 
V$_{\rm r}$=1000 km~s$^{-1}$ were used to normalize the spectra
in order to enable an easy comparison of the line shapes. Each subsequent
spectrum is shifted upwards by 0.5. The O I 8446 \AA\ line is also shown because of the
very similar system of narrow components (marked by short dashes)}
\end{center}
\label{f4}
\end{figure}

The presented spectroscopic behavior of V2275~Cyg will be compared with
other similar objects in Sect.\ 6, here we only mention that the spectral
properties correspond to a ``He/N'' nova in the classification system of
Williams (1992).

\begin{table}
\caption{Radial velocities for the narrow components in the O~I and
H$\alpha$ lines}
\begin{center}
\begin{tabular}{lclc}
\hline
Component & $V_r$ (km~s$^{-1}$) & Component & $V_r$ (km~s$^{-1}$)\\
\hline
a1 & $-$160 & e1 & $-$1250\\
a2 & 0      & e2 & $-$760\\
a3 & +360   & e3 & +1080\\
a4 & +760   & e4 & +1380\\
\hline
\end{tabular}
\end{center}
\end{table}

\section{The interstellar reddening}

Since the most interesting parameters of a nova system depend critically on
the inferred luminosity and distance, it is of paramount importance to
estimate the interstellar reddening and visual extinction. In this section,
we follow a similar discussion as in the case of CI~Aql (Kiss et al. 2001),
where similar observations resulted in an estimated $E(B-V)$. In this
study three different reddening determinations were applied.

First, we have surveyed all medium-resolution spectra to identify DIBs
from the list of Jenniskens \& D\'esert (1994). The only one DIB
that could be identified unambiguously is the one at 6613 \AA.
Although the sodium D spectrum covers the good 
reddening indicator DIB 5849 (Oudmaijer et al. 1997),
it was hidden in the higher noise level of the sodium D spectrum.
We measured the 
equivalent width (EW) of DIB 6613 with the IRAF task {\it splot} and 
found EW$_{6613}$=210$\pm$20 m\AA. This results in an
$E(B-V)=1\fm0\pm0\fm2$ (using the quoted uncertainty of the EW/$E(B-V)$
ratio in Jenniskens \& D\'esert 1994).

Second, two interstellar lines of Ca II 3933.66 and Na I D lines
were detected. The EW$_{\rm Ca II}$ can be converted into
EW$_{5780}$ with the empirical relationship 
EW$_{\rm Ca II}$/EW$_{5780}\approx$0.81 (Jenniskens \& D\'esert 1994). Therefore,
although we have not detected DIB 5780, we could estimate its strength from
the calcium line, yielding to an $E(B-V)=1\fm03\pm0\fm2$. The Na I D doublet
provides reliable excesses only in the moderately reddened regions (up to
$E(B-V)$=0\fm4, Munari \& Zwitter 1997), thus it is of lower significance.
The measured equivalent widths are EW$_{\rm NaD1}=1.07\pm0.02$ \AA\ and
EW$_{\rm NaD2}=0.92\pm0.02$ \AA. Their ratio is 1.16, being in good agreement
with the asymptotic behavior at high reddenings (Munari \& Zwitter 1997).
Therefore, the strong sodium doublet can only be used to place a lower 
limit on the reddening as $E(B-V)>0\fm7$ (see Fig.\ 2 in Munari \& Zwitter
1997).

Finally, the published $B-V$ colors can be also used to estimate the color
excess. The $(B-V)$ color of novae around maximum tends to be about
$B-V$=0\fm23$\pm$0\fm06 with a dispersion of 0\fm16 (Warner 1995). Two
magnitudes below maximum the dispersion decreases, therefore the
relation $(B-V)^{V(max)+2}_0\approx0\fm0$ can be used. As mentioned in the
Introduction, Sostero \& Leopardo measured $(B-V)$=1\fm1 and
1\fm0 at maximum light and three days later (without explicitly given
uncertainties). The resulting photometric reddening lies between 0\fm9 and
1\fm0. 

All of the reddening estimated scatter around 1\fm0. Despite the very good
agreement of the mean values (1\fm0, 1\fm03, 0\fm9, 1\fm0), there might be
higher systematic errors. In the following discussion we adopt the unweighted
mean of the spectroscopic and photometric reddenings 
$E(B-V)=1\fm0\pm0\fm1$ (estimated uncertainty). 

\section{The light curve}

\begin{figure}
\begin{center}
\leavevmode
\psfig{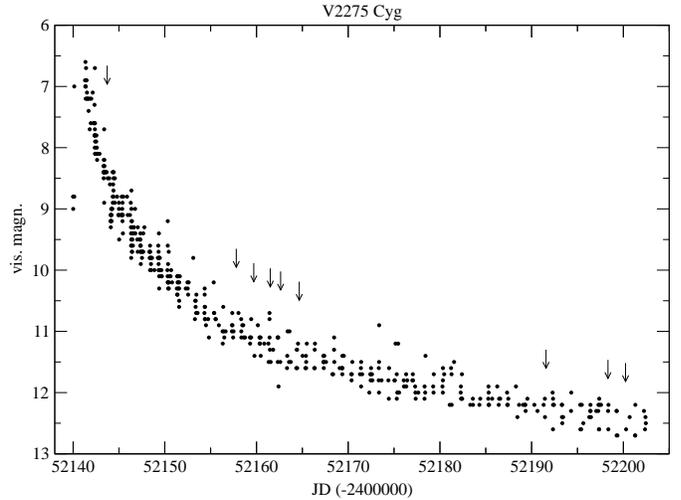}
\caption{The visual light curve of V2275~Cyg. The small arrows indicate
the epochs of spectroscopic observations}
\end{center}
\label{f5}
\end{figure}

In order to estimate the main photometric properties (epoch of maximum, 
rates of decline), we used the visual observations collected by the
VSNET group. The observational record consists of 494 individual visual 
estimates between the discovery and Oct. 19, 2001.

The light curve is plotted in Fig.\ 5. The epoch of maximum and decline 
rates were determined both from the original data and 0.1-day means.
Unfortunately, there are only few data around the maximum, thus, 
the inferred epoch, $t_0=2452141.4$ (2001 Aug. 19.9 UT) is uncertain 
by $^{+0.1}_{-0.5}$ d. The mean apparent maximum brightness was 6\fm8.
The fast decline is reflected by $t_2=2.9\pm0.5$ d
and $t_3=7\pm1$ d. Consequently, V2275~Cyg is a very fast nova. In fact,
Table\ 5.2 of Warner (1995) lists only 3 faster novae (V1500~Cyg, V838~Her and
MU~Ser). The epochs of spectroscopic observations are 
also marked in Fig.\ 5.

Three maximum magnitude versus rate of decline (MMRD) relations were used
to calculate visual absolute magnitude (Della Valle \& Livio 1995, 
Capaccioli et al. 1989, Schmidt 1957). They result in $-$9\fm4,
$-$10\fm6 and $-$9\fm0. The constant absolute magnitude 15 days after the
maximum (Capaccioli et al. 1989) gives $-$9\fm9. The unweighted average is
$M_{\rm V}=-9\fm7\pm0\fm7$ (formal error). The adopted reddening corresponds
to a total visual extinction $A_{\rm V}=3.1\times E(B-V)=3\fm1\pm0\fm3$.
Therefore, the estimated distance (using $m_{\rm max}=6\fm8$) 
$d=5^{+3}_{-2}$ kpc is quite inaccurate. A more meaningful
distance value needs proper modelling of the light curve 
in order to improve the calculated luminosity at maximum.

\section{Discussion}

Based on the presented spectroscopic behavior of V2275~Cyg we
identify the star as a member of the ``He/N'' subclass of 
classical novae following the definition by Williams (1992). All
of the defining characteristics are satisfied: broad lines (HWZI$>$
2500 km~s$^{-1}$) dominate the spectrum, the prominent lines 
are flat-topped with little absorption, a few forbidden lines 
occurred and $F({\rm He II 4686})\geq F({\rm H\beta})$. On the other hand, the 
very fast decline is also typical among the ``He/N'' novae.
According to Williams (1992), this means that the broader lines 
of the ``He/N'' emission spectrum originate in a discrete shell
ejected at high velocities from the white dwarf surface at
the peak of the outburst. The stable system of narrow components
in the H$\alpha$ and O I profiles support the presence of 
one or more discrete shells, as provided by the emission line
profile calculations of Gill \& O'Brian (1999). An 
alternative to the rings+polar caps geometry is inhomogeneous
ejecta with knots or clumps (see, e.g., Shore et al. 1997).

Since typical recurrent novae are classified as ``He/N'' novae, 
it is a natural consequence that the overall spectral appearance 
of V2275~Cyg is very similar to some well-observed recurrent novae.
The most recent example is CI~Aql (Kiss et al. 2001), where the
gross spectral characteristics are very similar. Furthermore, close
similarity is evident from a comparison with other recurrent novae,
such as U~Sco (Munari et al. 1999) or V394~CrA (Sekiguchi et al. 1989).
The question arises: could V2275~Cyg be a recurrent nova, observed
in outburst for the first time? We consider this to be quite unlikely. 
If the progenitor is indeed the star identified by Schmeer 
et al. (2001) with 18\fm8 USNO-A2.0 red magnitude, and our calculated 
absolute magnitude is approximately correct, then the resulting
absolute magnitude for the progenitor is M$_{\rm vis}\sim2-3$ mag.
This excludes the possibility of a red giant secondary as usual 
in recurrent novae. To fit the suggested progenitor with a typical 
red giant star, the outburst should have a visual absolute magnitude 
over $-$12 mag, much brighter than any nova outburst ever observed. 
If the progenitor was fainter and hidden by the suggested USNO-A2.0 star,
then the recurrent nova status is even less likely.

As expected from the very fast decline, V2275~Cyg closely resembles in some
aspects the well-studied fast nova, V1500~Cyg. Strittmatter et al.
(1977) presented a comparison of H$\alpha$ and O I 8446 profiles, where they
found a similar agreement between the line profiles to those
presented in Fig.\ 4. They concluded that the O I 8446 line is due to
Lyman $\beta$ fluorescence in clouds with high H$\alpha$ optical depth.
The strong resemblence suggest similar explanation in the
case of V2275~Cyg.

Further photometric as well as spectroscopic observations are expected
to extend the data baseline yielding to a better understanding of V2275~Cyg.
At present, neither the unambiguously identified progenitor,
nor the orbital period of the binary is known -- both are crucial 
for a reliable description of the system. On the other hand,
theoretical light curve modelling may place stronger constraints
on the luminosity of the outburst. 

\begin{acknowledgements}
This research was supported by the ``Bolyai J\'anos'' Research Scholarship
of LLK from the Hungarian Academy of Sciences, FKFP Grant 0010/2001,
Hungarian OTKA Grants \#T032258 and \#T034615, Pro Renovanda Cultura
Hungariae Grant DCS 2001. \'apr/6. and Szeged Observatory Foundation. The
NASA ADS Abstract Service was used to access data and references. We also
acknowledge the data service of the VSNET group. The authors express their
sincere thanks to Dr. M. De Robertis, who obtained one of the nova spectra
during his visit at DDO. This research has made use of Simbad Database
operated at CDS-Strasbourg, France.
\end{acknowledgements}

\end{document}